\documentclass[a4paper,11pt]{article}
\usepackage{pos}
\usepackage{lineno}

\title{Reconstruction in ALICE and calibration of TPC space-charge distortions in Run 3}

\manuallySeparateAuthors
\author*[a]{Ernst Hellb\"{a}r}
\author{for the ALICE collaboration}

\affiliation[a]{GSI - Helmholtzzentrum f\"{u}r Schwerionenforschung GmbH}

\emailAdd{ernst.hellbar@cern.ch}

\abstract{The ALICE experiment will run with continuous readout at interaction rates of up to 50 kHz in \mbox{Pb--Pb} collisions during Run 3 of the LHC. In order to achieve this goal, a new data processing scheme and software are developed. This scheme strongly relies on GPUs (Graphics Processing Unit) for fast online and offline calibration and reconstruction as well as on efficient data compression. On the hardware side, the Time Projection Chamber (TPC), among other detector systems, received  major upgrades to its readout chambers and readout electronics. The multiwire proportional chambers were replaced by stacks of four Gas Electron Multiplier foils to allow for continuous readout while keeping the ion backflow below 1\%, minimizing space-charge effects from amplification ions entering the drift volume. Nevertheless, significant space-point distortions due to space charge are expected at the highest interaction rates in \mbox{Pb--Pb} collisions. In addition, space-charge density fluctuations lead to distortion fluctuations which have to be corrected on time scales of 10 ms in order to preserve the intrinsic tracking resolution of the TPC. While the average space-charge distortions can be corrected using information from external detectors as a reference, data-driven machine learning algorithms and convolutional neural networks are foreseen to provide the correction for the distortion fluctuations.}

\FullConference{%

 The Ninth Annual Conference on Large Hadron Collider Physics - LHCP2021\\
 7-12 June 2021\\
 Online
}

\begin{document}
\maketitle

\section{Introduction}
The ALICE experiment aims to record a Pb--Pb minimum-bias data sample in Run 3 and Run 4 of the LHC which is 50--100 times larger than that recorded in the first two running periods. This will be realized by fully exploiting the capabilities of the LHC, which will deliver Pb--Pb collisions at interaction rates of up to 50 kHz. Contrary to the event-based triggered readout which was employed before Run 3, ALICE will now continuously process and record the data provided by the detectors, i.e. data from every collision will be available. Continuous readout poses several challenges and new requirements for the global data processing strategy \cite{O2TDR}, including the calibration and reconstruction software and, for some of the detector systems, also requiring hardware upgrades. The data processing is performed in two stages, \textit{synchronous processing} and \textit{asynchronous processing}. The former takes place during data taking and includes a first calibration and tracking, while the final calibration and full reconstruction of the data is performed during the \textit{asynchronous processing} at a later point in time. At both stages, the application of GPUs for a majority of the processing leads to a significant gain in computing time, which is required to continuously process the amounts of data expected to be delivered in Run 3 and Run 4.
\\ \\
The Time Projection Chamber (TPC) \cite{ALICETPC} is the main detector for tracking and particle identification in ALICE. During the long shutdown of the LHC before Run 3, the readout chambers and electronics were upgraded \cite{UpgradePaper} in order to provide the capability to run in continuous readout mode. Instead of multiwire proportional chambers and a triggered gating grid, which effectively limited the readout rate to about 3 kHz in order to avoid space-charge effects from ion backflow (IBF), the upgraded readout chambers are based on Gas Electron Multiplier (GEM) foils. A combination of four GEM foils with different hole pitches and an optimized high-voltage configuration suppresses the IBF to below 1\%, eliminating the need for a gating grid and allowing the operation of the detector with continuous readout. Taking into account the high interaction rates foreseen for Pb--Pb collisions in Run 3, significant space-point distortions due to space charge from IBF are still expected, requiring a sophisticated calibration approach to preserve the intrinsic tracking performance of the TPC.


\section{Space-charge effects in the TPC}
\begin{figure}[t]
  \centering
  \includegraphics[width=0.47\textwidth]{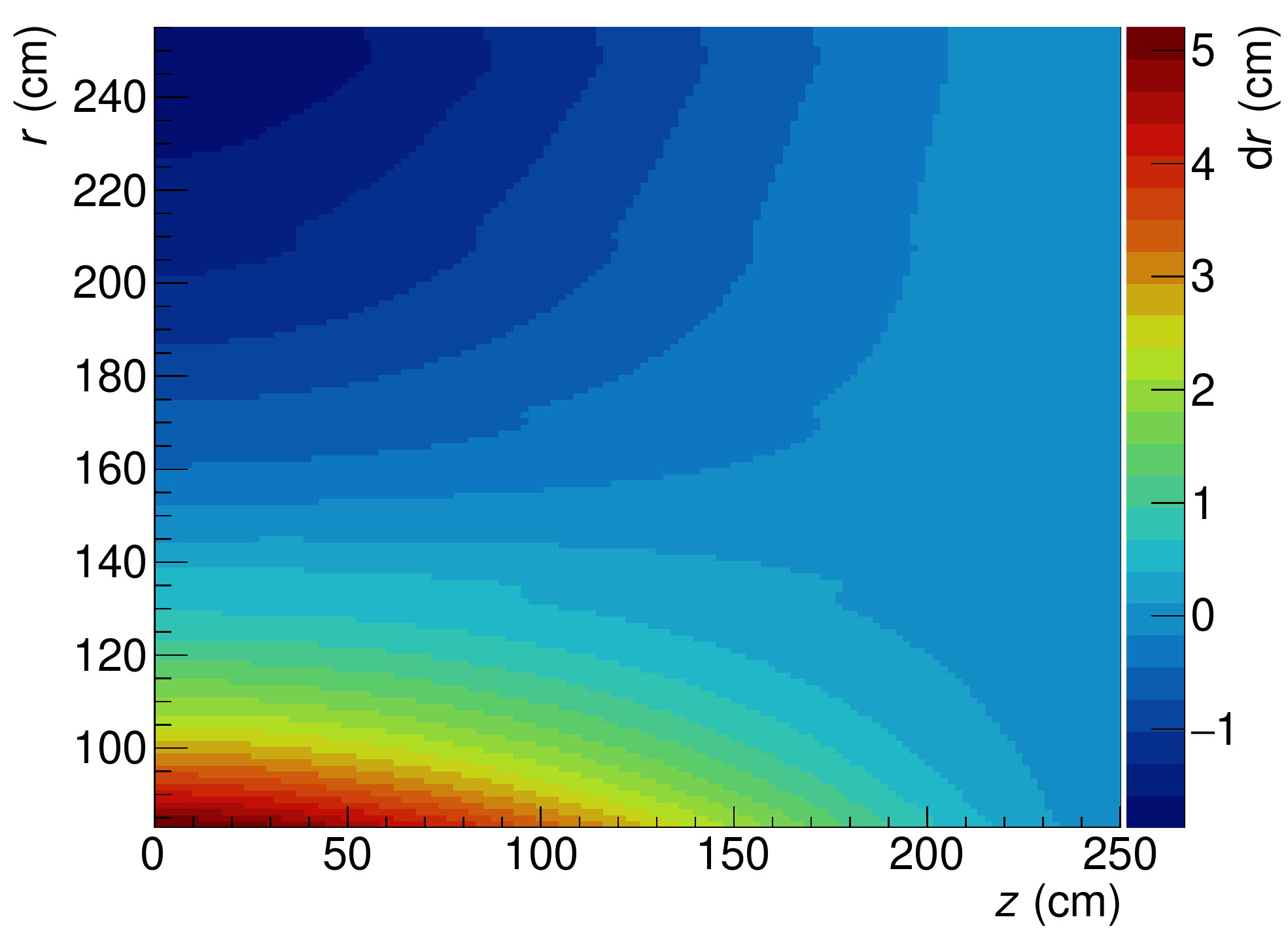}
  \caption{The radial space-charge distortions $\mathrm{d}r$ (colour axis) as a function of the radius $r$ and the $z$-position for Pb--Pb collisions at 50 kHz and $\varepsilon \approx 10$.}
  \label{fig:TDRDist}
\end{figure}

\begin{figure}[t]
  \centering
  \includegraphics[width=0.77\textwidth]{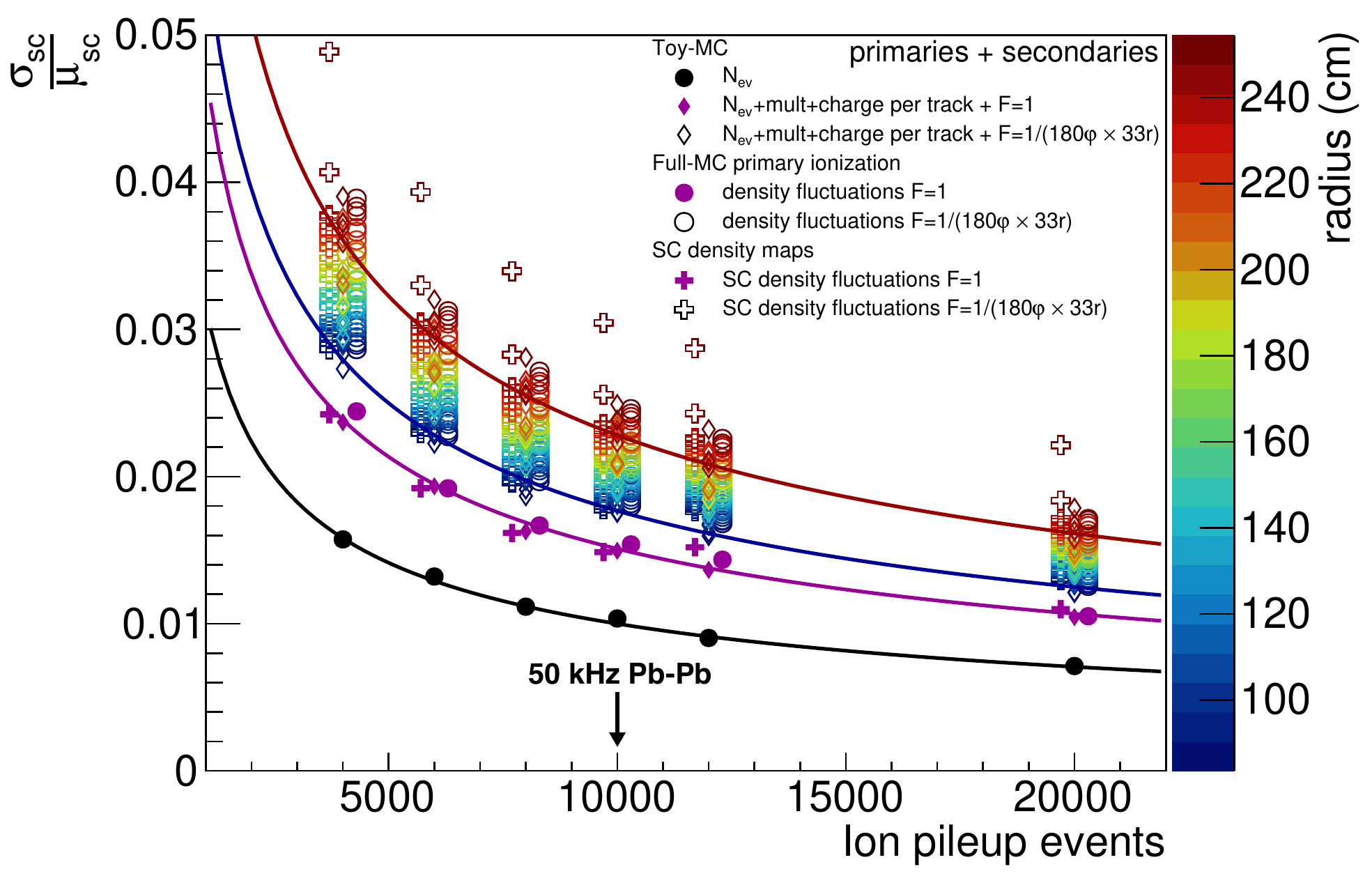}
  \caption{The relative space-charge density fluctuations $\frac{\sigma_{\mathrm{SC}}}{\mu_{\mathrm{SC}}}$ as a function of the number of ion pile-up events, estimated from equation~\ref{eq:fluc} (lines), toy MC simulations (diamonds), full MC simulations (circles) and simulations based on space-charge density maps (crosses). The black markers and line represent only the contribution from the fluctuations of the number of ion pile-up events. The purple markers and line show the relative fluctuations of all contributions integrated over the full TPC volume ($F = 1$). The open markers in rainbow colours illustrate the radial dependence of the fluctuations when dividing the TPC volume into small volume elements, e.g. here into 180 bins in $\varphi$ and 33 bins in $r$ ($F = 1 / (180\varphi \times 33r)$). The blue and red line represent the fluctuations obtained from equation~\ref{eq:fluc} at the innermost and outermost radius, respectively, for the same volume elements.}
  \label{fig:SCFluc}
\end{figure}

The space-charge density depends on the amount of ions piling up within one ion drift time.  It is driven by the number of ion pile-up events, which is the product of the collision rate and the ion drift time, as well as by the effective gain and the IBF, which are combined to $\varepsilon = \mathrm{gain} \times \mathrm{IBF}$. The expected space-charge density distribution at 50 kHz Pb--Pb is derived using the hit distributions from Monte Carlo simulations. As dedicated measurements of the ion mobility in Ne-CO$_{2}$-N$_{2}$ (90-10-5) \cite{Deisting} indicate an ion drift time around 200 ms for the full drift, ions from 10000 events pile up on average in the TPC drift volume at any given moment at 50 kHz interaction rate. The space-charge distortions $\mathrm{d}r$, $\mathrm{d}r\varphi$ and $\mathrm{d}z$ for any given space-charge density distribution are calculated using numerical algorithms to solve the Poisson equation and the Langevin equation. The average radial space-charge distortions for 50 kHz Pb--Pb collisions are shown in Fig.~\ref{fig:TDRDist} as a function of the TPC radius and $z$-position.
\\ \\
The space-charge density fluctuations $\sigma_{\mathrm{SC}}$ relative to the average space-charge density $\mu_{\mathrm{SC}}$ can be calculated analytically by
\begin{equation}
  \label{eq:fluc}
  \frac{\sigma_{\mathrm{SC}}}{\mu_{\mathrm{SC}}} = \frac{1}{\sqrt{N^{\mathrm{ion}}_{\mathrm{pile-up}}}} \sqrt{1 + \left( \frac{\sigma_{N_{\mathrm{mult,tot}}}}{\mu_{N_{\mathrm{mult,tot}}}} \right) ^2 + \frac{1}{F_{\mu_{\mathrm{tot}}}(F, r)} \left( 1 + \left( \frac{\sigma_{Q_{\mathrm{track,tot}}}}{\mu_{Q_{\mathrm{track,tot}}}}  (r) \right) ^2 \right) } \; ,
\end{equation}
where $N^{\mathrm{ion}}_{\mathrm{pile-up}}$ is the mean number of ion pile-up events, $\left(\frac{\sigma_{N_{\mathrm{mult,tot}}}}{\mu_{N_{\mathrm{mult,tot}}}} \right) ^2$ is the relative RMS of the distribution of the track multiplicity, $\mu_{N_{\mathrm{mult}}}$ is the average track multiplicity per event and $\left( \frac{\sigma_{Q_{\mathrm{track,tot}}}}{\mu_{Q_{\mathrm{track,tot}}}}  (r) \right) ^2$ is the relative variation of the ionization of single tracks depending on the \mbox{radius $r$}. The track-based quantities are composed of contributions from primary (prim) and secondary (sec) tracks. The function $F_{\mu_{\mathrm{tot}}}(F, r)  = F_{\mathrm{prim}}(F, r)\cdot\mu_{N_{\mathrm{mult,prim}}} + F_{\mathrm{sec}}(F, r)\cdot\mu_{N_{\mathrm{mult,sec}}}$ quantifies the amount of tracks contributing to the fluctuations for a given volume fraction $F$, i.e. $F = 1$ for the full TPC volume and $F < 1$ for a fraction of the volume. The relative space-charge density fluctuations are shown as a function of the number of ion pile-up events in Fig.~\ref{fig:SCFluc}, using both equation~\ref{eq:fluc} and MC simulations. They are plotted for the full TPC volume as well as for only a fraction of the volume as a function of the radius. For smaller volume elements and towards outer radii, the relative fluctuations are generally larger as fewer tracks contribute.  Relative space-charge density fluctuations of the order of 2\% are expected at 50 kHz interaction rate, leading to distortion fluctuations of the order of several millimeters. It is demonstrated in \cite{UpgradeTDR} that the distortion fluctuations need to be corrected in time intervals of the order of 10 ms to preserve the intrinsic tracking resolution of the TPC of \mbox{200 $\mu$m}.

\section{Calibration of space-charge distortions and distortion fluctuations}
The correction of the average space-charge distortions and the distortion fluctuations is performed in multiple steps during both \textit{synchronous} and \textit{asynchronous reconstruction}, aiming to restore the intrinsic track resolution of the TPC after the final calibration. The algorithm for the average space-charge distortion correction is outlined in \cite{Schmidt}. The external detectors ITS (Inner Tracking System), TRD (Transition Radiation Detector) and TOF (Time-Of-Flight detector) are used to extract a 3D correction map from the data. The calibration interval for the average correction is of the order of minutes in order to collect enough statistics to reach an accuracy of the average-correction map of the order of \mbox{50 $\mu$m}. As the relevant time scales for the distortion fluctuations (10 ms) are much shorter than the calibration interval of the average correction, an additional correction for the fluctuations is required. Fast data-driven machine learning (ML) algorithms and convolutional neural networks (CNN) are developed to predict the corrections for the distortion fluctuations. The distortion fluctuations depend both on the space-charge density fluctuations and on the average space-charge density, requiring both of these quantities as input for the ML and CNN models. The space-charge density fluctuations are derived from the integrated digital currents (IDC) which are the charge signals on each TPC readout pad integrated over \mbox{1 ms}. Several pads are grouped in pad and row direction and the integrated charge is averaged to obtain the final 3D IDCs as a function of radius ($r$), azimuth ($\varphi$) and time (corresponding to the $z$-coordinate). Integrating the 3D IDCs over $r$ and $\varphi$, the 1D IDCs as a function of time are obtained, still containing information about the fluctuations of the number of ion pile-up events and of the track multiplicity. The average space-charge density is estimated from the numerical derivative of the average corrections w.r.t. the IDCs. It is extracted from data using the ITS-TRD-TOF method and the Fourier coefficients of the 1D IDC Fourier transform for small time windows within the average calibration interval. Two types of corrections are foreseen for the distortion fluctuations. A 1D$\rightarrow$3D correction provides a 3-dimensional correction of part of the fluctuations, using the Fourier coefficients of the 1D IDC Fourier transform and the derivative of the average correction as input. Simple ML models based on boosted decision trees are expected to perform well. While this type of correction will be sufficient for pp collisions, additionally a 3D$\rightarrow$3D correction will be applied for Pb--Pb collisions. A CNN based on the U-Net architecture \cite{Unet} will be employed, using the 3D IDCs and the derivative of the average correction as input variables.

\section{Summary}
ALICE switched from a triggered to a continuous readout scheme for Run 3 of the LHC, requiring the development of a new data processing scheme and software as well as several upgrades of the detector systems. The correction of space-charge distortions in the upgraded TPC is one of the most challenging tasks of the calibration. The average space-charge distortions are corrected in time intervals of about 1 min by extracting correction maps from the data using the external detectors ITS, TRD and TOF as references. The correction of the distortion fluctuations needs to be updated on time scales of the order of 10 ms. It will be performed using trained ML and CNN models to predict the corrections from the 1D or 3D IDCs and the derivative of the average corrections.

\end{document}